\documentclass[lettersize,journal]{IEEEtran}
\usepackage{amsmath,amsfonts}
\usepackage{algorithmic}
\usepackage{algorithm}
\usepackage{array}
\usepackage[caption=false,font=normalsize,labelfont=sf,textfont=sf]{subfig}
\usepackage{textcomp}
\usepackage{stfloats}
\usepackage{url}
\usepackage{verbatim}
\usepackage{graphicx}
\usepackage{cite}
\usepackage{xcolor}
\usepackage{enumitem}
\usepackage{hyperref}
\usepackage{amsmath,amssymb,amsfonts}
\usepackage{orcidlink}

\begin{document}
\title{User Satisfaction-Aware Resource Allocation with Prospect-Theoretic Utility for Multimedia Streaming}
\author{
\begin{tabular}{ccc}
Manoj Kumar S$^{*}$ & Pranjal Varshney$^{*}$ & Avhishek Chatterjee
\end{tabular}
\thanks{The authors are with the Department of Electrical Engineering, Indian Institute of Technology Madras, Chennai, Tamil Nadu 600036, India. E-mails: {ee24d020,ee23b180}@smail.iitm.ac.in, avhishek@ee.iitm.ac.in. \\
$^{*}$Student authors are listed alphabetically. }
}

\markboth{TO be submitted in IEEE Journal ,~Vol.~xx, No.~x, XX~2026}%
{Shell \MakeLowercase{\textit{et al.}}: A Sample Article Using IEEEtran.cls for IEEE Journals}

\IEEEpubid{}
\maketitle
\begin{abstract}
Traditional resource allocation policies for multimedia streaming have primarily targeted metrics such as throughput, fairness and delay. However, user satisfaction is known to be strongly influenced by variations in throughput. Although some recent works \cite{VinayGustavo_01, Th_var_01, Th_var_02} have considered throughput variations, they overlooked a well-known aspect of human behavior that has reshaped behavioral economics over the past few decades \cite{Prospect_P1}: the same amount of decrease in throughput causes greater annoyance than the satisfaction caused by the increase. Motivated by this insight from the broad field of economics, popularly known as prospect theory, we propose a generic utility (metric) that captures both the average throughput and the asymmetric effect of throughput variations on user satisfaction. Next, we propose a resource allocation policy that reduces abrupt decreases in allocated resources when the system transitions from a resource-rich to resource-scarce state while respecting the priorities of different classes of users. Extensive simulations show that this policy outperforms the existing fair allocation policies. However, the choice of the optimal threshold depends on the network instance, which, in practice, may change frequently. Hence, building on the theoretical analysis of multi-queue systems, we propose fast numerical methods for determining the optimal threshold, which can be used in field deployments.
\end{abstract}

\begin{IEEEkeywords}
Resource allocation; Prospect theory; Queuing analysis. 
\end{IEEEkeywords}

\section{Introduction}
\label{sec:Introduction}

A characteristic feature of human behavior is loss aversion, as noted in the Nobel-winning work on behavioral economics \cite{Prospect_P1}. In particular, humans are known to experience losses more intensely than the satisfaction derived from equivalent gains. Current quality of experience (QoE) metrics for  multimedia streaming largely overlook this inherent asymmetry in human behavior. Specifically, most QoE metrics consider average metrics such as data rate, buffering rate or average delay. Although there is a vast body of literature that studies variations in data rate, the QoE metrics considered therein are symmetric, i.e., the same magnitude of increase and decrease in data rate (delay) is treated identically. 

In this work, we introduce a general QoE metric that captures this asymmetry by building on prospect theory \cite{Prospect_P1}. We propose a threshold-based resource allocation policy that achieves better QoE than the fair allocation policy, which maximizes the total average throughput. Furthermore, using an analytical approach, we propose computationally efficient methods for obtaining the optimal threshold value, since the system settings change frequently due to user dynamics and other changes in the network. Our proposed method and the corresponding analysis naturally extends to networks with multiple classes of users.

\subsection{Related Literature}

Resource allocation is fundamental to wireless communication systems, where limited radio resources, such as bandwidth, time slots, power, etc. must be efficiently distributed among users. In traditional resource allocation methods, such as maximum throughput (MT) \cite{MT_paper}, proportional fair (PF) \cite{PF_paper}, and modified largest weighted delay first (M-LWDF) \cite{MLWDF_paper}, 
the objectives are to optimize throughput or delay, or to enforce  fairness. However, user satisfaction in multimedia streaming is governed by the frequency of buffering pauses, multimedia quality and temporal variations of the multimedia quality \cite{Userengagement_P2,Ref_02_Other_01,Ref_02_Other_02,Akhil_01}.

A significant amount of research has been devoted to QoE-aware resource allocation, where scheduling decisions are guided by utility functions that map network-level parameters such as throughput, delay, packet loss and bitrate, to user-perceived quality. In \cite{QoE_Scheduler_P4, QoE_Scheduler_P5, Ref_04_Other_01, Ref_04_Other_02,TCOM_Ref_04}, cross-layer resource allocation frameworks jointly optimize application-layer video adaptation and network-layer resource allocation to maximize QoE, thereby improving resource utilization compared with the conventional throughput-based schemes. In \cite{QoE_Scheduler_P6, Ref_06_Other_01}, the authors proposed QoE-aware scheduling frameworks that employ learning-based QoE estimation, QoE-aware prioritization and resource allocation based on channel conditions, buffer status and video requirements to reduce playback interruptions while achieving a balance among QoE, throughput and fairness. 

In \cite{QoE_Scheduler_P7, Ref_07_Other_01,Akhil_01}, buffer-aware and QoE-aware resource allocation strategies were proposed to improve playback continuity in adaptive video streaming. By incorporating playback buffer status and bitrate requirements into resource allocation decisions, these approaches reduce playback stalls and enhance overall user-perceived quality. In \cite{QoE_Scheduler_P8,Ref_08_Other_01, TCOM_REf_08}, packet importance, delay sensitivity, video quality metrics and video characteristics are incorporated into resource allocation to improve resource utilization while preserving video quality. In \cite{Userengagement_P3}, a control-theoretic framework that formulates bitrate adaptation as a stochastic optimization problem is proposed to improve user QoE for HTTP-based adaptive video streaming services, with QoE explicitly defined as a weighted combination of average bitrate, quality variations, rebuffering time and startup delay. 
There is a line of work \cite{VinayGustavo_01, Th_var_01, Th_var_02} that captured QoE by introducing utility functions that increase with average quality and decrease with variations in quality, and proposes resource allocation algorithms to improve QoE by optimizing these utility functions or by provisioning constrained rate variability while efficiently utilizing network resources.

Furthermore, several multi-class resource allocation algorithms have been proposed to address heterogeneous user requirements by assigning different priorities to different user classes. In \cite{Multiclass_Scheduler_P9, TCOM_Ref_09}, a class-based priority scheduling and cross-layer resource allocation schemes assigns priority weights to different user classes and exploit application-layer information, thereby influencing resource allocation along with channel conditions. Similarly, \cite{Multiclass_Scheduler_P10} proposed a service-based prioritization approach that preferentially allocates resources to QoE-sensitive video streaming users while distributing the remaining resources among lower-priority services. In \cite{Multiclass_Scheduler_P11, Multiclass_Scheduler_P12} QoE fairness-based schemes that prioritize users with poor QoE to ensure a baseline level satisfaction while improving overall system performance. 

\subsection{Main contribution}
\label{sec:Main_contribution}

As discussed above, a well known concept in behavioral economics \cite{Prospect_P1} is that a drop in quality is perceived more intensely than an equivalent improvement in quality. Existing QoE metrics and related utility functions, including those that capture variations in quality \cite{VinayGustavo_01, Th_var_01, Th_var_02}, consider the impact of such variations to be symmetric. Furthermore, in most QoE-aware algorithms for networks with multiple classes of users, the number of users in each class is assumed to remain static throughout the resource allocation process. To address these limitations, we make the following main contributions.

\begin{enumerate}[label=\roman*)]
    \item We propose a performance metric that captures the asymmetric impact of variations in average quality or data rate. The  metric is defined as a linear combination of an average resource allocation metric, which reflects the average amount of resources received by users, and a prospect metric, which captures the asymmetric impact of increase and decrease in the allocated resources. 

    \item Next, we propose a novel threshold-based resource allocation policy that allocates a stable share of resources when the number of active users is below a  threshold, thereby avoiding unnecessary reductions in allocation due to minor changes in the system load. Once the number of users exceeds this threshold, the available resources are distributed equally among all users. Thus, the proposed policy helps reduce abrupt decreases in resource allocation, which are perceived more negatively by users. 

    \item An $M/M/\infty$ system model was developed to evaluate the performance of the proposed resource allocation algorithm. The simulation results demonstrate that the proposed threshold-based resource allocation policy achieves better performance than the existing policies. 

    \item Due to the stochastic nature of user arrivals and departures, determining the optimal threshold through simulation requires repeated evaluations over a wide range of threshold values, resulting in high execution time. Therefore, we develop an exact  theoretical framework based on numerical analysis, that accurately computes the optimal threshold approximately 100 times faster than the simulation.

    \item Furthermore, we introduce an approximate analytical approach derived from the exact theoretical framework by considering a single bystander that remains in the system at all times. This simplifies the analysis while preserving the system's essential characteristics. Numerical results show that the bystander approach achieves performance metrics and optimal threshold value similar to the exact theoretical analysis, with approximately 30 times faster execution. These computational advantages make the proposed analytical approaches well suited for efficient system design and optimization.

\end{enumerate}

We emphasize that our resource allocation operates at the time scale of user dynamics rather than at the time-scale of channel coherence times. For  wireless systems with $n$ users and $m$ fading channels (subcarriers), due to concentration of probability, each user $1\le i \le n$ can be allocated $k_i$ channels at the highest fading state, i.e., the highest permissible modulation rate, provided that $\sum_{i=1}^n k_i \le m$, when $m \sim 10^2$, which is the case for orthogonal frequency-division multiple access (OFDMA) in 4G/5G systems. We refer to \cite{Akhil_01} for a detailed proof of this result, particularly the proof of Theorem~3 and the accompanying discussions.  Hence, we do not consider the effects of fading and adaptive modulation and coding (AMC) in this work, and focus on  user dynamics.

\subsection{Organization of the paper}

The rest of the paper is organized as follows. Section~\ref{sec:Problem_setting} presents the problem formulation including user dynamics, user utility modeling and different user classes. The proposed threshold-based resource allocation policy and weighted threshold-based resource allocation policy for multi-class scenarios, along with their comparison with processor sharing (PS)/ fair allocation (FA) policy is presented in section~\ref{sec:Resource_Allocation_Policy}.  Section~\ref{sec:Exact_Theoretical_Analysis} presents the exact theoretical framework and simulation-based validation for both single-class and double-class systems. Section~\ref{sec:Bystander_approach} introduces an approximate analytical framework based on the bystander approach and evaluates its performance for single-class and double-class scenarios. The comparative performance evaluation of the exact theoretical and  bystander approaches is provided in section~\ref{sec:comparative_performance_evaluation}. Finally, the paper concludes with section~\ref{sec:Conclusion}.

\section{Problem setting}
\label{sec:Problem_setting}

\subsection*{User Dynamics}

The system is modeled as an $M/M/\infty$ queue. User arrivals follow a Poisson process with rate 
$\lambda$ and occur independently and randomly over time. Each user requires service for a random duration that is exponentially distributed with rate $\mu$. The service times are assumed to be independent across users and independent of the arrival process. The system has an infinite number of servers, ensuring that every arriving user immediately receives service without any waiting. The total available resource in the system is denoted by $B>0$.

Let $\rho = \lambda / \mu$ denote the traffic intensity of the system. In the steady state, the number of users present in the system is a random variable that follows a Poisson distribution with mean $\rho$. The stationary probability that the system contains exactly $k$ users is given by
\begin{equation}\label{eq:pi_k}
\pi(k) = e^{-\rho}\frac{\rho^k}{k!}, \qquad k = 0,1,2,\ldots \nonumber
\end{equation}

\subsection*{User Utility}
The performance metric used for optimization is the average amount of resources allocated to a user, which is denoted by $A(F)$, where $F$ is the resource allocation policy.  Let $T$ denote the duration for which a user remains in the system. Then, the average amount of resources allocated is given by
\begin{equation}
A(F) = \frac{1}{T}\int_{0}^{T} F\bigl(q(t)\bigr)\, dt
\end{equation}
where $q(t)$ denotes the number of users in the system at time $t$ and $F(q)$ denotes the amount of resources allocated to a particular user. When there are $q$ users in the system, the resource allocation is symmetric and each person in the system is allocated $F(q)$ amount of resources.

User behavior is also modeled based on prospect theory using two constants, $\alpha$ and $\beta$, where $\alpha > \beta$. An increase of $x$ units of resources adds $\beta x$ to the metric, while a decrease of $x$ units of resources subtracts $\alpha x$.

Let $s_i$ $(i = 1, \ldots, k_1)$ represent all time instants at when a user's allocated resources increase by an amount of $g_{i} \, (i = 1, \ldots, k_1)$ and let $t_i$ $(i = 1, \ldots, k_2)$ represent all time instants at which the user’s allocated resources decrease by an amount of $h_i$ $(i = 1, \ldots, k_2)$. The prospect-based metric $P(F)$, is given by
\begin{equation}
P(F) = \beta \sum_{i=1}^{k_1} g_{i} - \alpha \sum_{i=1}^{k_2} h_i
\end{equation}

The overall performance metric $M(F)$ combines the
average resources metric $A(f)$ and the prospect-based metric $P(F)$, weighted by a factor $\gamma$ and is given by
\begin{equation}
M(F) = A(F) + \gamma P(F)
\end{equation}

\subsection*{User Class}
In general, the system may serve users belonging to multiple classes that differ in their traffic intensities and in how they perceive the services. Each class has its own arrival rate, indicating how frequently users of that class enter the system. The service time is modeled as a common parameter, implying that the session durations of users across all classes are statistically identical. Furthermore, the prospect constants are also class dependent. Users belonging to different classes exhibit different degrees of gain and loss sensitivity, represented by separate parameters that quantify the extent to which they perceive resource reductions negatively and value resource increases. 

This framework allows the system to be modeled with two distinct classes, namely class-1 and class-2, with $\lambda_{c_1}$ and $\lambda_{c_2}$ denoting their respective arrival rates and $\mu$ denoting the common service rate. The arrival rate for class-2 is set higher than that of class-1, i.e.  $\lambda_{c_1} < \lambda_{c_2}$. This reflects the assumption that class-2 users have higher arrival intensity and impose a heavier overall load on the system than class-1 users. The prospect gain and loss constants for class-1 and class-2 are denoted as $\beta_{c_1}$, $\beta_{c_2}$, $\alpha_{c_1}$ and $\alpha_{c_2}$, respectively. The system is modeled under the assumption that the class-1 users are more sensitive to slight degradation in service than class-2 users. Hence, the prospect loss constants satisfy $\alpha_{c_1} > \alpha_{c_2}$, whereas the prospect gain constants satisfy $\beta_{c_1} < \beta_{c_2}$. The overall performance metric is obtained by aggregating the contributions of both classes.

\section{Resource Allocation Policy}
\label{sec:Resource_Allocation_Policy}
Resource allocation is the process of distributing a fixed amount of resources among all active users in the system to optimize performance, fairness and service quality. In shared systems, the demand for resources fluctuates as users join or leave the system. Therefore, a resource allocation policy is required to determine each user's share and maximize resource utilization. 

As discussed in Sec.~\ref{sec:Main_contribution}, thanks to previous work \cite{Akhil_01} that established concentration of measure for user-to-subcarrier mapping  in multichannel wireless systems with fading and AMC, e.g. OFDMA in 4G/5G. Consequently, it is sufficient to study resource allocation at the scale of user dynamics instead of fading.

\subsection*{Processor Sharing (PS) / Fair Allocation (FA) Policy \cite{PS_book}} This is the simplest resource allocation policy, in which the total available resources are equally shared among all active users in the system. Let $q$ denote the number of active users in the system and $B$ denote the total available resources. Then, the resource allocation based on PS/FA policy is given by

\begin{align}
    N(q) = \frac{B}{q}
\end{align}
This policy is conceptually straightforward and is commonly used in real-world systems because it is easy to implement and automatically adapts to the system load. However, this policy ignores the user's sensitivity to quality fluctuations. When the number of users in the system is small, a new arrival can cause a large drop in the resources allocated to all the active users, which may be perceived as highly negative under prospect-theory. In other words, this policy can lead to a large and sudden reduction in the service experienced by existing users as the population increases.

\subsection*{Threshold-based Resource Allocation Policy}
To mitigate sudden fluctuations in service, a new threshold-based resource allocation policy is proposed. When there are $q$ users in the system, the amount of resources allocated to each user under the proposed threshold-based resource allocation is given by

\begin{align}
    F(q)=
      \begin{cases}
        \dfrac{B}{t}, & q \le t\\[6pt]
        \dfrac{B}{q}, & q > t
      \end{cases}
\end{align}

where $B$ denotes the total available resource and $t$ is the threshold. This policy intentionally limits the resource allocated under low system load so that the arrival of a new user does not cause a large and sudden drop in the resources allocated to the existing users. This is more beneficial in terms of the prospect metric, under which users are more sensitive to losses than to equivalent gains. The main objective of this policy is to determine the threshold $t$ that maximizes the overall performance metric.

\subsection*{Weighted Threshold-based Resource Allocation Policy for multi-class systems} 
As the system can accommodate users belonging to multiple classes, the proposed approach is extended to a weighted threshold-based resource allocation policy. Each class is assigned a positive weight and the system monitors the weighted load rather than the actual the user count. For a system with $N$ classes, the resource allocation to the tagged user belonging to class-$i$, where $i \in N$ is given by

\begin{align}
    G_{c_i}(q_i)=
      \begin{cases}
        \dfrac{w_{c_i}B}{t}, & S_i(\mathbf{q}) \le t\\[6pt]
        \dfrac{w_{c_i}B}{S_i(\mathbf{q})}, & S_i(\mathbf{q}) > t
      \end{cases}
\end{align}
where $\mathbf{q} = \left[q_{c_1}, q_{c_2}, \cdots, q_{c_n}   \right]$ is the vector representing the number of users in each class and $S_i$ denotes the total weighted load, which is given by 

\begin{align}
    S_i(\mathbf{q}) = \sum_{n \, \in N} w_{c_n} q_{c_n} \, +\, w_{c_i}     \nonumber
\end{align}

 Here, $w_{c_n}$ is the weight associated with class-$n$ and $w_{c_i}$ is the weight associated with the class to which the tagged user belongs.

Taking into account a system with two classes, namely class-1 and class-2, having weights $w_{c_1}\, \text{and}\, w_{c_2}$, respectively, the resource allocation to a tagged user according to the proposed weighted threshold-based resource allocation policy is

\begin{align}
    G_{c_i}(q_{c_1}, q_{c_2})=
      \begin{cases}
        \dfrac{w_{c_i}B}{t}, & S_i(q_{c_1},q_{c_2}) \le t\\[6pt]
        \dfrac{w_{c_i}B}{S_i(q_{c_1},q_{c_2})}, & S_i(q_{c_1},q_{c_2}) > t
      \end{cases}
\end{align}

The total weighted load in this case is given by 
\begin{align}
    S_i(q_{c_1},q_{c_2}) =  w_{c_1} q_{c_1} \,+\, w_{c_2} q_{c_2} \,+\, w_{c_i} \nonumber
\end{align}

By applying the threshold to the total weighted load and allocating resources proportionally, this policy ensures stable per-user allocation under low system load and fair resource sharing among heterogeneous classes under high system load. The main objective of this policy is to determine the optimal class weights and threshold that jointly maximize the performance metric.

\subsection*{Comparative analysis with PS/FA resource allocation policy}

A comparative analysis is conducted between the proposed threshold-based resource allocation policy for single-class, the proposed weighted threshold-based resource allocation policy for double-class systems, and the existing PS/FA resource allocation policy. The simulations are carried out using the same experimental setup and performance metrics to ensure a fair comparison. 

\subsubsection*{Single-class}
\begin{figure}[htbp]
\centerline{\includegraphics[scale = 0.5]{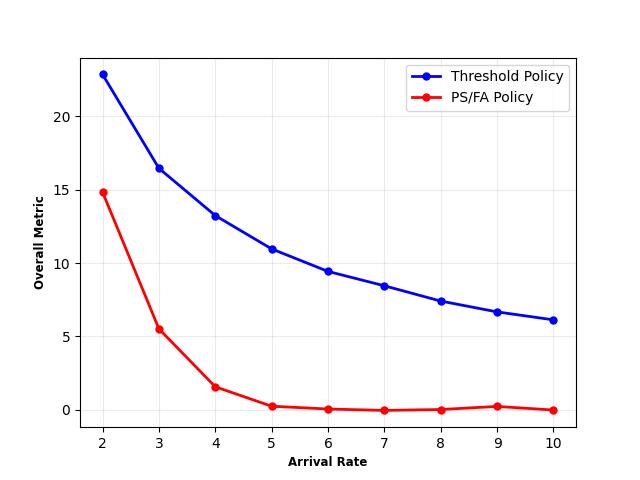}}
\caption{Comparison between PS/FA and the proposed threshold-based resource allocation policy for varying arrival rates}
\label{fig:Fig_11_Allocation_policy_Singlepack_Comp}
\end{figure}

Fig.~\ref{fig:Fig_11_Allocation_policy_Singlepack_Comp} shows the variation of the overall performance metric $M[f]$ for varying arrival rates $\lambda$ under the PS/FA and threshold-based resource allocation policies. As the arrival rate increases, the overall performance metric decreases for both policies, indicating that the system performance degrades with increasing traffic load and reduced resource allocation per user. However, the threshold-based allocation policy consistently achieves significantly higher metric values than the PS/FA allocation policy across different arrival rates. Under the PS/FA resource allocation policy, frequent fluctuations in the allocated resources occur whenever the number of users in the system changes. This has a greater impact on the prospect metric, resulting in a lower performance metric than that achieved by the threshold-based resource allocation policy. This demonstrates that the threshold-based resource allocation policy allocates the available resource more efficiently.

\subsubsection*{Double-class} 
A similar comparison is carried out for the double-class system by comparing the proposed weighted threshold-based allocation policy with  PS/FA policy. Under the PS/FA policy, resources are allocated to the two classes in proportion to their arrival rates, $\lambda_{c_1} : \lambda_{c_2}$ and each arriving user can access only the portion assigned to their respective class. Within each class, the allocated resources are shared equally among all users present in the system. This policy therefore serves as a simple baseline without thresholds or weighting mechanisms.

Due to the varying arrival rates, the prospect sensitivity is allowed to vary with the relative scarcity of the classes. Let $r$ denote the arrival rate ratio
\begin{equation}
    r=\frac{\lambda_{c_2}}{\lambda_{c_1}} \ge 1 
    \nonumber
\end{equation}

The class-specific parameters are scaled as
\begin{align}
\alpha_{c_1} &= \alpha_0\ r^{a} & \alpha_{c_2} &= \alpha_0\, r^{-a} \notag \\ 
\beta_{c_1}  &= \beta_0\, r^{-b} & \beta_{c_2}  &= \beta_0\, r^{b} 
\nonumber
\end{align}

where $\alpha_0>\beta_0>0$ are base parameters and $0<a<b<1$. 

The intuition is that when $\lambda_{c_2}$ greatly exceeds $\lambda_{c_1}$, reductions in allocated resources are perceived more strongly by class-1 users, leading to greater loss sensitivity and weaker response to gains. Conversely, class-2 users  becomes less sensitive to losses and more responsive to incremental gains. The proposed scaling therefore captures this behavior while maintaining the structural condition $\alpha_{c_i} > \beta_{c_i}$ for each class.

For each parameter configuration, the expected performance metric under the PS/FA policy, $\mathbb{E}[M(N)]$ and the expected performance metric under the weighted threshold-based resource allocation policy, $\mathbb{E}[M(G)]$ are evaluated. The performance is summarized using the ratio between the two performance metric obtained using the two allocation policies, given by
\begin{equation}
    \eta = \frac{\mathbb{E}[M(N)]}{\mathbb{E}[M(G)]}
    \nonumber
\end{equation}

 \begin{figure}[htbp]
\centerline{\includegraphics[scale = 0.5]{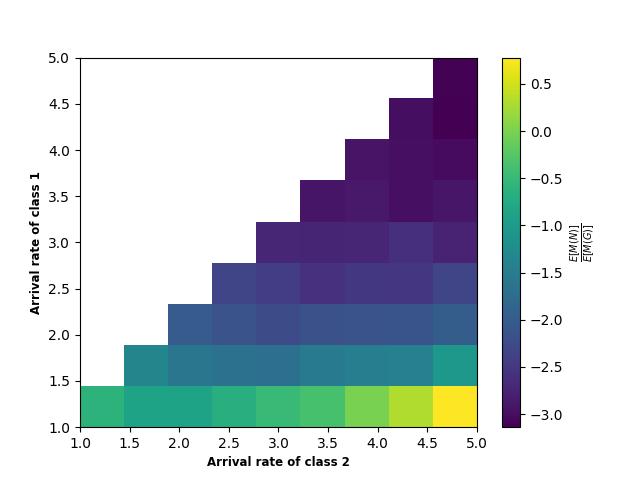}}
\caption{Heatmap of the ratio of the performance metrics for different combinations of the arrival rates of class-1 and class-2 users}
\label{fig:Fig_12_Varying_lambdas_policies_comparision}
\end{figure}

Figure~\ref{fig:Fig_12_Varying_lambdas_policies_comparision} shows that, across the entire parameter grid the ratio remains strictly below $1$, indicating that the PS/FA resource allocation policy never outperforms the weighted threshold-based policy in terms of the performance metric $E[M]$. Furthermore, the ratio is negative, because the prospect component dominates the allocation term under the PS/FA policy, causing the overall metric $\mathbb{E}[M(N)]$ fall below zero. In contrast, the weighted threshold-based policy consistently maintains significantly higher values of $\mathbb{E}[M(G)]$ demonstrating a clear advantage.

Another notable pattern is observed along the horizontal slices of the heatmap. For a fixed value of $\lambda_{c_2}$, the ratio exhibits a uni-modal concave profile as $\lambda_{c_1}$ varies. This suggests that the relative performance gap between the two policies varies non-monotonically with the arrival-rate imbalance between the classes.

\subsection*{Finding the best threshold and weights in real time}

The above comparisons shows that our threshold-based resource allocation policy achieve significantly better performance. However, to achieve the best performance, the threshold $t$ in the single-class case and the threshold $t$ and the weights $w_{c_1}$ and $w_{c_2}$ in the double-class case, have to be chosen optimally. For the above plots, these parameters were determined through exhaustive search using simulations.

In practical field deployments, performing extensive simulations for every network setting is computationally challenging. To address this, analytical expressions are derived using Markov analysis and fast numerical methods for finding the best threshold and weights are developed based on these analytical expressions. The derived analytical expressions are formulated as a system of linear equations and the computation of the performance metrics involves solving this system of linear equations. Detailed analytical derivations are presented in the subsequent sections.

\section{Exact Analysis}
\label{sec:Exact_Theoretical_Analysis}
In the considered system, users enter and depart from the system independently over time, resulting in a dynamically varying set of active users at any given instant. Rather than focusing on an individual user, this approach considered the collective behavior of all the active users in the system. The performance evaluation is done by averaging the resources allocated across all the users, thereby offering an aggregate view of system behavior. 

\subsection*{Theoretical Analysis}
Using the underlying stochastic model with Markov chain and probabilistic methods, the theoretical analysis aims to derive a closed-form expression for the overall performance metric.

\subsubsection*{Single-Class}
In the single-class system, the theoretical analysis focuses on the rewards received by a typical (tagged) user conditioning on the number of other $q$ users present in the system at the time of arrival. By the Poisson arrival see time averages (PASTA) property, the distribution seen by an arriving user is the same as the stationary distribution of the system, i.e. $\pi(q)$ is the steady state probability of having $q$ other users in the system, which is also the probability that a randomly chosen arriving user observes $q$ other users in the system. 

A new event corresponds to the arrival of a new user or the departure of another user or the departure of the tagged user from the system. The probabilities of these events are given by  

\begin{equation}\label{eq:event_probs}
P_+(q)=\frac{\lambda}{\Lambda(q)}\quad P_-(q)=\frac{q\mu}{\Lambda(q)}\quad P_{\mathrm{Leave}}(q)=\frac{\mu}{\Lambda(q)}
\nonumber
\end{equation}

where $P_+(q)$, $P_-(q)$ and $ P_{\mathrm{Leave}}(q)$ are the probability of the arrival of a new user or the departure of another user or the departure of the tagged user from the system respectively, and satisfy $P_+(q) + P_-(q) +  P_{\mathrm{Leave}}(q) = 1$. The waiting time until the next event is exponentially distributed with rate $\Lambda(q)$ and is given by

\begin{equation}\label{eq:Lambda_q}
\Lambda(q)=\lambda + q\mu + \mu = \lambda + (q+1)\mu
\end{equation}

One of the main quantities of interest is the expected cumulative resources $\mathbb{E}[R_D(q)]$ received by the tagged user from the time of arrival until departure, conditioned on there being $q$ other user being in the system. The derivation of $\mathbb{E}[R_D(q)]$ is given by 

\begin{align}
\mathbb{E}[R_D(q)] &={} 
P_+(q)\,\, \mathbb{E}[R_D(q)\mid  q+1] \notag\\
& + \,P_-(q)\,\, \mathbb{E}[R_D(q)\mid  q-1]  \notag\\
& + \,P_{\,\text{Leave}}(q)\, \,  \mathbb{E}[R_D(q)\mid q_{\,\text{Leave}}]
\nonumber
\end{align}
Applying the event probabilities and conditional expectations,

\begin{align}
\mathbb{E}[R_D(q)] &={} 
\frac{\lambda}{\Lambda(q)} \left( \frac{F(q+1)}{\Lambda(q)} + \mathbb{E}[R_D(q+1)] \right)  \notag\\
& + \,\frac{q\mu}{\Lambda(q)} \left( \frac{F(q+1)}{\Lambda(q)} + \mathbb{E}[R_D(q-1)] \right)  \notag\\
& + \,\frac{\mu}{\Lambda(q)} \left(\frac{F(q+1)}{\Lambda(q)} \right)
\nonumber
\end{align}
This can be further simplified as
\begin{align}
\mathbb{E}[R_D(q)] &={} 
 \frac{F(q+1)}{\Lambda(q)}   \notag\\
& + \frac{\lambda}{\Lambda(q)} \, \mathbb{E}[R_D(q+1)]  \notag\\ 
& + \,\frac{q\mu}{\Lambda(q)} \, \mathbb{E}[R_D(q-1)]   
\end{align}

By the Renewal Reward Theorem (RRT), the long-run average reward rate is given by the ratio of the expected cumulative resource received during a renewal cycle to the expected cycle length. Therefore, the expected time-averaged resource $\mathbb{E}[R_A(q)]$ conditioned on the tagged user observing $q$ other users upon arrival arrival is given by

\begin{align}
    \mathbb{E}[R_A(q)]  = \frac{1}{\mu} \mathbb{E}[R_D(q)]
\end{align}

The other main quantity is the expected cumulative prospect contribution $\mathbb{E}[R_P(q)]$ that captures the gains and losses in the resources allocated to the tagged user resulting from the arrival or departure of another user in the system. The quantity $\mathbb{E}[R_P(q)]$ is given by

\begin{align}
\mathbb{E}[R_P(q)] &={} 
P_+(q)\,\, \mathbb{E}[R_P(q)\mid  q+1 ] \notag\\
& + \,P_-(q)\,\, \mathbb{E}[R_P(q)\mid  q-1 ]  
\nonumber
\end{align}

\begin{align}
\mathbb{E}[R_P(q)] &={} 
\frac{\lambda}{\Lambda(q)} \Big(-\alpha\big(F(q+1)-F(q+2)\big)  \Big) \notag\\
&  + \, \frac{q\mu}{\Lambda(q)} \Big(\beta\big(F(q)-F(q+1)\big)  \Big) \notag\\
&  + \, \frac{\lambda}{\Lambda(q)} \Big( \mathbb{E}[R_P(q+1)] \Big) \notag\\
& + \frac{q\mu}{\Lambda(q)} \Big( \mathbb{E}[R_P(q-1)]\Big)
\end{align}

Using the PASTA property, the expected time-averaged resource metric and the expected prospect metric are given by 

\begin{align}
    \mathbb{E}[A(F)] = \sum_{q = 0}^{\infty} \, \pi(q) \, \mathbb{E}[R_A(q)] 
\end{align}

\begin{align}
    \mathbb{E}[P(F)] = \sum_{q = 0}^{\infty} \, \pi(q) \, \mathbb{E}[R_P(q)] 
\end{align}

Finally, the theoretical expression for the overall performance metric $\mathbb{E}[M(F)]$ under the threshold-based resource allocation policy is given by 

\begin{align}
    \mathbb{E}[M(F)] = \mathbb{E}[A(F)] \, + \, \gamma \, \mathbb{E}[P(F)] 
\end{align}

\subsubsection*{Double-class}
Theoretical analysis is carried  out by considering the arrival of tagged user from either of the two classes with varying number of users in each class class. Let $q_{c_1}$ and $q_{c_2}$ denote the number of class-1 and class-2 users respectively. By the PASTA property, the joint distribution of ($q_{c_1}, q_{c_2}$) observed at the arrival instant coincides with the stationary distribution of the underlying Markov chain. Since the users of both classes are modeled as independent Poisson random variables, the probability that the tagged user finds $q_{c_1}$ users of class-1 and $q_{c_2}$ users of class-2 in the system is 

\begin{align}
    \Pi(q_{c_1}, q_{c_2}) = \pi(q_1) \, \pi(q_{c_2})
    \nonumber
\end{align}

The total rate of ($q_{c_1}, q_{c_2}$) with respect to class-1 arrival, class-2 arrival, departure of other class-1 user, departure of other class-2 user and departure of the tagged user is given by

\begin{align}
    \Lambda(q_{c_1},q_{c_2}) = \lambda_{c_1} + \lambda_{c_2} + (q_{c_1} + q_{c_2} + 1)\mu 
    \nonumber
\end{align}

The event probabilities are given by 

\begin{align}
    P_{+_{c_1}} (q_{c_1}, q_{c_2}) &= \frac{\lambda_{c_1}}{\Lambda(q_{c_1}, q_{c_2})} \quad P_{+_{c_2}} (q_{c_1}, q_{c_2}) = \frac{\lambda_{c_2}}{\Lambda(q_{c_1}, q_{c_2})} \notag \\
    P_{-_{c_1}} (q_{c_1}, q_{c_2}) &= \frac{q_{c_1} \mu}{\Lambda(q_{c_1}, q_{c_2})} \quad P_{-_{c_2}} (q_{c_1}, q_{c_2}) = \frac{q_{c_2} \mu}{\Lambda(q_{c_1}, q_{c_2})} \notag \\
    P_{\text{Leave}} (q_{c_1}, q_{c_2}) &= \frac{\mu}{\Lambda(q_{c_1}, q_{c_2})}
    \nonumber
\end{align}

For a tagged user of class $i \in\{ 1,2\}$,  $\mathbb{E}[R^{(c_i)}_D (q_{c_1}, q_{c_2})]$ denotes the expected cumulative resources acquired by the tagged user from arrival until departure, conditioned on the initial state ($q_{c_1}, q_{c_2}$). The recursion for $\mathbb{E}[R^{(c_i)}_D (q_{c_1}, q_{c_2})]$ is derived as

\begin{align}
    \mathbb{E}[R^{(c_i)}_D (q_{c_1}, q_{c_2})] &={}   P_{+_{c_1}} (q_{c_1}, q_{c_2}) \,\,\mathbb{E}[R^{(c_i)}_D (q_{c_1}, q_{c_2})\,|\,  q_{c_1} + 1] \notag \\
    & + P_{+_{c_2}} (q_{c_1}, q_{c_2}) \,\,\mathbb{E}[R^{(c_i)}_D (q_{c_1}, q_{c_2})\,|\,  q_{c_2} + 1] \notag \\
    & + P_{-_{c_1}} (q_{c_1}, q_{c_2}) \,\, \mathbb{E}[R^{(c_i)}_D (q_{c_1}, q_{c_2})\,|\,  q_{c_1} - 1] \notag \\
    & + P_{-_{c_2}} (q_{c_1}, q_{c_2}) \,\, \mathbb{E}[R^{(c_i)}_D (q_{c_1}, q_{c_2})\,|\,  q_{c_2} - 1] \notag \\
    & + P_{\,\text{Leave}} (q_{c_1}, q_{c_2}) \,\,\mathbb{E}[R^{(c_i)}_D (q_{c_1}, q_{c_2})\,|\, q^{(c_i)}_{\,\text{Leave}}]
    \nonumber
\end{align}

\begin{align}
    &\mathbb{E}[R^{(c_i)}_D (q_{c_1}, q_{c_2})] ={} \frac{\mu}{\Lambda(q_{c_1}, q_{c_2})} \frac{ G_{c_i}(q_{c_1}, q_{c_2})}{\Lambda(q_{c_1}, q_{c_2})} \notag \\&+  \frac{\lambda_{c_1}}{\Lambda(q_{c_1}, q_{c_2})} \,\, \left(\mathbb{E}[R^{(c_i)}_D (q_{c_1}+1, q_{c_2})] +  \frac{G_{c_i}(q_{c_1}, q_{c_2})}{\Lambda(q_{c_1}, q_{c_2})}\right)  \notag \\ 
    & + \frac{\lambda_{c_2}}{\Lambda(q_{c_1}, q_{c_2})} \,\, \left(\mathbb{E}[R^{(c_i)}_D (q_{c_1}, q_{c_2}+1)] + \frac{G_{c_i}(q_{c_1}, q_{c_2})}{\Lambda(q_{c_1}, q_{c_2})}\right) \notag \\ 
    & + \frac{q_{c_1} \mu}{\Lambda(q_{c_1}, q_{c_2})} \,\,\left(\mathbb{E}[R^{(c_i)}_D (q_{c_1}-1, q_{c_2})] + \frac{G_{c_i}(q_{c_1}, q_{c_2})}{\Lambda(q_{c_1}, q_{c_2})} \right) \notag \\ 
    & + \frac{q_{c_2} \mu}{\Lambda(q_{c_1}, q_{c_2})} \,\, \left(\mathbb{E}[R^{(c_i)}_D (q_{c_1}, q_{c_2}-1)] + \frac{G_{c_i}(q_{c_1}, q_{c_2})}{\Lambda(q_{c_1}, q_{c_2})}\right)\notag
    \nonumber
\end{align}

\begin{align}
    \mathbb{E}[R^{(c_i)}_D (q_{c_1}, q_{c_2})] &={}   \frac{G_{c_i}(q_{c_1}, q_{c_2})}{\Lambda(q_{c_1}, q_{c_2})}  \notag \\ 
    & + \frac{\lambda_{c_1}}{\Lambda(q_{c_1}, q_{c_2})} \,\, \mathbb{E}[R^{(c_i)}_D (q_{c_1}+1, q_{c_2})] \notag \\  
    & + \frac{\lambda_{c_2}}{\Lambda(q_{c_1}, q_{c_2})} \,\, \mathbb{E}[R^{(c_i)}_D (q_{c_1}, q_{c_2}+1)] \notag \\ 
    & + \frac{q_{c_1} \mu}{\Lambda(q_{c_1}, q_{c_2})} \,\,\mathbb{E}[R^{(c_i)}_D (q_{c_1}-1, q_{c_2})] \notag \\
    & + \frac{q_{c_2} \mu}{\Lambda(q_{c_1}, q_{c_2})} \,\, \mathbb{E}[R^{(c_i)}_D (q_{c_1}, q_{c_2}-1)] 
\end{align}

The expected time-averaged resource for class $i$ is given by 
\begin{align}
    \mathbb{E}[R^{(c_i)}_A(q_{c_1}, q_{c_2})]  = \frac{1}{\mu} 
    \, \mathbb{E}[R^{(c_i)}_D(q_{c_1}, q_{c_2})]
\end{align}

Similarly, for a tagged user of class $i \in\{ 1,2\}$,  $\mathbb{E}[R^{(c_i)}_P (q_{c_1}, q_{c_2})]$ denotes the expected cumulative prospect contribution received by the tagged user from arrival until departure conditioned on the initial state ($q_{c_1}, q_{c_2}$). The recursion for $\mathbb{E}[R^{(c_i)}_P (q_{c_1}, q_{c_2})]$ is derived as

\begin{align}
    \mathbb{E}[R^{(c_i)}_P (q_{c_1}, q_{c_2})] &={}   P_{+_{c_1}} (q_{c_1}, q_{c_2}) \,\,\mathbb{E}[R^{(c_i)}_P (q_{c_1}, q_{c_2})\,|\, q_{c_1} + 1] \notag \\
    & + P_{+_{c_2}} (q_{c_1}, q_{c_2}) \,\,\mathbb{E}[R^{(c_i)}_P (q_{c_1}, q_{c_2})\,|\,  q_{c_2} + 1] \notag \\
    & + P_{-_{c_1}} (q_{c_1}, q_{c_2}) \,\, \mathbb{E}[R^{(c_i)}_P (q_{c_1}, q_{c_2})\,|\,  q_{c_1} - 1] \notag \\
    & + P_{-_{c_2}} (q_{c_1}, q_{c_2}) \,\, \mathbb{E}[R^{(c_i)}_P (q_{c_1}, q_{c_2})\,|\,  q_{c_2} - 1] 
    \nonumber
\end{align}

\begin{align}
     \mathbb{E}[R^{(c_i)}_P (q_{c_1}, q_{c_2})] &={}
     \frac{\lambda_{c_1}}{\Lambda(q_{c_1}, q_{c_2})} \Big(\mathbb{E}[R^{(c_i)}_P (q_{c_1}+1, q_{c_2})]  \Big) \notag \\
     & + \frac{\lambda_{c_1}}{\Lambda(q_{c_1}, q_{c_2})} \Big( -\alpha_{c_i} \, \Delta(q_{c_1}+1, q_{c_2}) \Big) \notag\\
     & + \frac{\lambda_{c_2}}{\Lambda(q_{c_1}, q_{c_2})} \Big(\mathbb{E}[R^{(c_i)}_P (q_{c_1}, q_{c_2}+1)]  \Big) \notag \\
     & + \frac{\lambda_{c_2}}{\Lambda(q_{c_1}, q_{c_2})} \Big( -\alpha_{c_i} \, \Delta(q_{c_1}, q_{c_2}+1) \Big) \notag\\
     & + \frac{q_{c_1} \mu}{\Lambda(q_{c_1}, q_{c_2})} \Big(\mathbb{E}[R^{(c_i)}_P (q_{c_1}-1, q_{c_2})] \Big) \notag \\
     & + \frac{q_{c_1} \mu}{\Lambda(q_{c_1}, q_{c_2})} \Big( \beta_{c_i} \, \Delta(q_{c_1}-1, q_{c_2})  \Big) \notag \\
     & + \frac{q_{c_2} \mu}{\Lambda(q_{c_1}, q_{c_2})} \Big(\mathbb{E}[R^{(c_i)}_P (q_{c_1}, q_{c_2}-1)] \Big) \notag \\
     & + \frac{q_{c_2} \mu}{\Lambda(q_{c_1}, q_{c_2})} \Big( \beta_{c_i}\, \Delta(q_{c_1}, q_{c_2}-1)\Big) 
\end{align}
where $\Delta(q_{c_1}+1, q_{c_2})$ and $\Delta(q_{c_1}, q_{c_2}+1)$ are the decrease in the allocated resources when a class-1 user and a  class-2 user joins the system, respectively. Similarly, $\Delta(q_{c_1}-1, q_{c_2})$ and $\Delta(q_{c_1}, q_{c_2}-1)$ denote the increase in the resource when a class-1 user and a class-2 user departs from the system, respectively.

\begin{align}
    & \Delta(q_{c_1}+1, q_{c_2}) = G^{(c_i)}(q_{c_1}, q_{c_2})-G^{(c_i)}(q_{c_1}+1,q_{c_2}) \notag \\
    & \Delta(q_{c_1}, q_{c_2}+1) = G^{(c_i)}(q_{c_1}, q_{c_2})-G^{(c_i)}(q_{c_1},q_{c_2}+1) \notag \\
    & \Delta(q_{c_1}-1, q_{c_2}) = G^{(c_i)}(q_{c_1}-1,q_{c_2})-G^{(c_i)}(q_{c_1},q_{c_2}) \notag \\
    & \Delta(q_{c_1}, q_{c_2}-1) =  G^{(c_i)}(q_{c_1},q_{c_2}-1)-G^{(c_i)}(q_{c_1},q_{c_2})
    \nonumber
\end{align}

The expected time-averaged resource metric and the expected prospect metric are given by

\begin{align}
    \mathbb{E}[A^{(c_i)}(G)] = \sum_{q_{c_1}, q_{c_2}} \, \Pi(q_{c_1}, q_{c_2}) \, \mathbb{E}[R^{(c_i)}_A(q_{c_1}, q_{c_2})] 
\end{align}

\begin{align}
    \mathbb{E}[P^{(c_i)}(G)] = \sum_{q_{c_1}, q_{c_2}} \, \Pi(q_{c_1}, q_{c_2}) \, \mathbb{E}[R^{(c_i)}_P(q_{c_1}, q_{c_2})] 
\end{align}

The class-wise overall metric is 
\begin{align}
    \mathbb{E}[M^{(c_i)}(G)] = \mathbb{E}[A^{(c_i)}(G)] \, + \, \gamma \, \mathbb{E}[P^{(c_i)}(G)] 
\end{align}

Since a randomly chosen arrival belongs to class-1 with probability $\lambda_{c_1}/(\lambda_{c_1} + \lambda_{c_2})$ and to class-2 with probability $\lambda_{c_2}/(\lambda_{c_1} + \lambda_{c_2})$, the combined overall performance metric is given by

\begin{align}
    \mathbb{E}[M(G)] &= \frac{\lambda_{c_1}}{\lambda_{c_1} + \lambda_{c_2}}\, \mathbb{E}[M^{(c_1)}(G)] \notag \\
     & + \frac{\lambda_{c_2}}{\lambda_{c_1} + \lambda_{c_2}}\, \mathbb{E}[M^{(c_2)}(G)]
\end{align}

\subsection*{Validation through Simulation Analysis}
Simulation-based evaluation involves the numerical simulation of the $M/M/\infty$ system by randomly generating arrival and departure events and analyzing the user experience throughout the user's stay in the system. The simulation provides empirical estimates of the expected average resource, expected prospect contribution and overall performance metric for a range of threshold values.

The simulation is organized as a sequence of discrete events. At each event, an arrival or a departure occurs randomly, with the time between successive events drawn from an exponential distribution. Between consecutive events, all the users in the system receive a constant resource allocation determined by threshold-based policy, which is accumulated into each user's cumulative resource. When arrival or departure occurs, the number of users in the system increases or decreases respectively and the corresponding prospect loss or gain is added to all the active users in the system. The time-averaged resource and the prospect contribution of the departing users are used to determine the overall performance metric. To eliminate transient effects, an initial set of  departing users is discarded and the simulation is continued until either a sufficient number of users have departed or the maximum number of events has occurred.

\subsubsection*{Single-class}
\begin{figure}[htbp]
\centerline{\includegraphics[scale = 0.5]{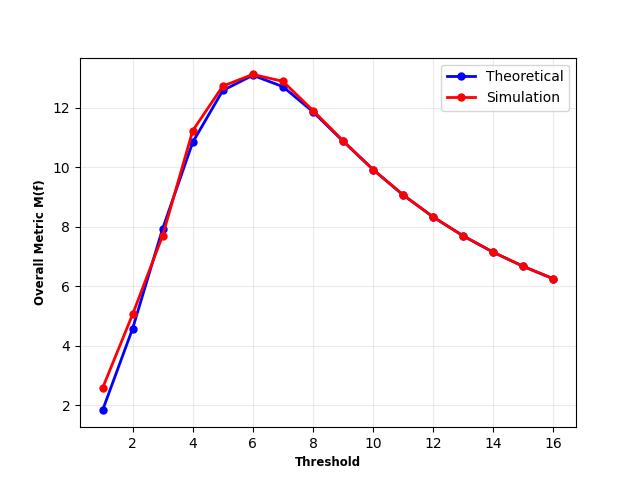}}
\caption{Overall performance metric comparison obtained through \\ simulation and exact theoretical analysis for single-class scenario}
\label{fig:Fig_01_Avg_single}
\end{figure}

For single-class simulation analysis, the $M/M/\infty$ system is configured with an arrival rate $\lambda$ = 4, a service rate $\mu$ = 1 and a total available resources of $B$ = 100. The  Prospect loss sensitivity and gain sensitivity constants are set to $\alpha$ = 2 and $\beta$ = 1, respectively. The scalar weight $\gamma$ = 1 is used to combine the average resource and the prospect contribution. To obtain a reliable steady-state estimates, the first 500 departing users are discarded. The maximum number of departing users  and events considered in the simulation are 20,000 and 50,000,000 respectively. The overall performance metric is evaluated over a threshold range of 1 to 16. The comparison of the overall performance metric obtained from the simulation and the theoretical analysis for varying threshold value is shown in Fig.~\ref{fig:Fig_01_Avg_single}.

The variation of the overall performance metric with the threshold illustrates the trade-off introduced by the threshold-based resource allocation policy. When the threshold is very low, the available resource is divided more aggressively. As a result, frequent reductions in the allocated resources occur whenever a new user arrives. As the threshold increases, the system becomes more tolerant to the arrival of new users. This enables a more balanced allocation of the resource, resulting in a steady increase in the overall performance metric. The metric reaches its maximum at the threshold $t = 6$, indicating that this threshold provides an efficient and balanced allocation of resources with minimal reductions  caused by fluctuations in the allocated resources. Beyond this point, the metric gradually decreases as the system allocates resources more conservatively when the number of users exceeds the threshold. Consequently, the per-user resource allocation decreases significantly, thereby reducing the overall performance metric. In addition, the close agreement between the theoretical and  simulation curves over entire threshold values validates the correctness of the analytical model and confirms that the theoretical formulation accurately captures the system behavior.

\subsubsection*{ Double-class}

The system is modeled with two classes, class-1 and class-2 with arrival rate $\lambda_{c_1} = 2$ and $\lambda_{c_2} = 6$ respectively, common service parameter $\mu = 1$ and total available resource of $B = 100$. The prospect loss constants are set to $\alpha_{c_1} = 4$ and $\alpha_{c_2} = 3$, while the prospect gain constants are set to $\beta_{c_1} = 1$ and $\beta_{c_2} = 2$. The scalar weight used to compute the overall metric is set to $\gamma = 1$. Similar to the single-class simulation, the first 500 departing users are discarded and the maximum number of departing users and events considered are 20,000 and 50,000,000 respectively. The weight of class-1 is varied as $w_{c_1} \in \{0.5, 1,2,3,4,5 \}$, while the weight of class-2 is kept constant at $w_{c_2} = 1$. The simulation is carried out over a threshold values range of 1 to 30 and the overall performance metric is evaluated.

\begin{figure}[htbp]
\centerline{\includegraphics[scale = 0.55]{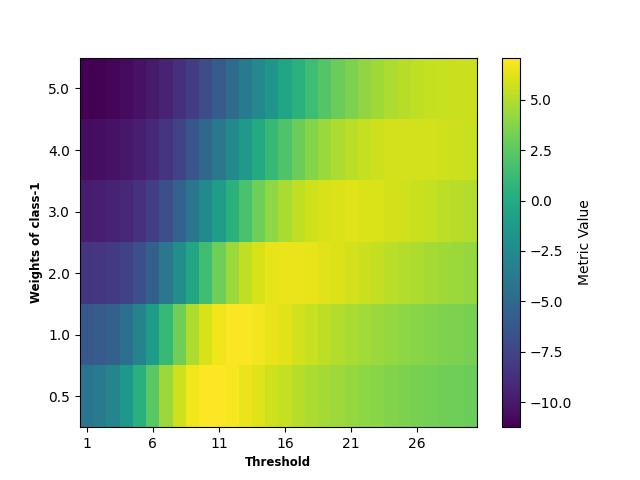}}
\caption{Heatmap of performance metric across threshold  and class-1 \\ weight values under exact theoretical analysis}
\label{fig:Fig_02_Avg_Double_Th_Heatmap}
\end{figure}

\begin{figure}[htbp]
\centerline{\includegraphics[scale = 0.55]{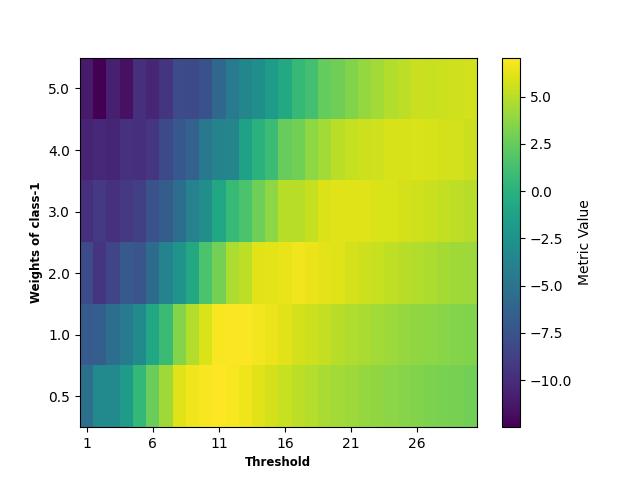}}
\caption{Heatmap of performance metric across threshold  and class-1 \\ weight values under simulation analysis}
\label{fig:Fig_03_Avg_Double_Sim_Heatmap}
\end{figure}

The heatmap in Fig.~\ref{fig:Fig_02_Avg_Double_Th_Heatmap} and Fig.~\ref{fig:Fig_03_Avg_Double_Sim_Heatmap} depicts how the overall performance metric for users from two different classes varies with different class weights and threshold values under theoretical and simulation analyses, respectively. From the figures, it can be observed that, for each value of $w_{c_1}$, the metric initially increases as threshold increases and reaches a maximum.  At very small threshold values, the metric value is low, as indicated by the darker shades in the heatmaps. This suggests that the resource allocation is highly sensitive to the instantaneous number of users in the system, resulting in less efficient resource distribution. As the threshold increases, the metric gradually increases, as indicated by the transition from darker to brighter shades in the heatmaps. This increase in the overall metric indicates reduced fluctuations in the resources allocated among users. As a result, users in both the class receive more consistent service. 

Another important observation from the heatmaps is that the highest metric value shifts towards larger threshold values as the weight increases. As $w_{c_1}$ increases, the weighted demand contributed by the class-1 users also increases, effectively increasing the weighted load in the system. To maintain a balanced resource allocation, a larger threshold is required. This explains the diagonal pattern of the high metric region observed in the heatmaps. Furthermore, beyond the optimal threshold, increasing the threshold value allocates the resources more uniformly without providing additional benefits. Consequently, the overall metric decreases significantly. The heatmaps clearly demonstrate the importance of properly tuning the threshold values for different weights to achieve optimal system performance. 

\begin{figure}[htbp]
\centerline{\includegraphics[scale = 0.5]{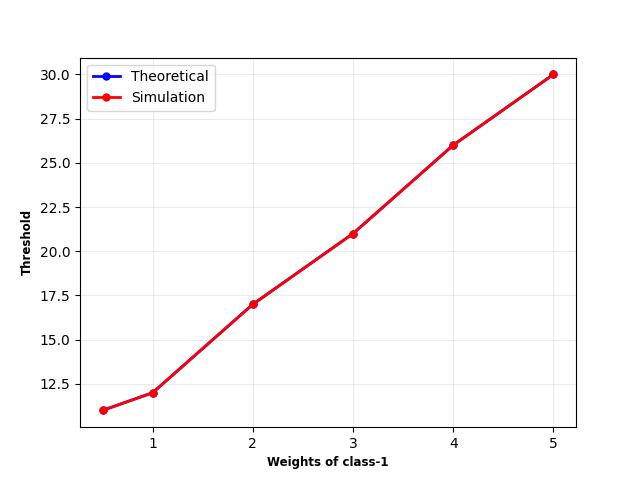}}
\caption{Comparison of optimal threshold for different class-1 weights \\ under theoretical and simulation analysis}
\label{fig:Fig_04_Avg_Double_Threshold_Comp}
\end{figure}

Figure~\ref{fig:Fig_04_Avg_Double_Threshold_Comp} shows the comparison of the optimal threshold for different class-1 weights obtained under the theoretical and simulation analyses. From the figure, it is evident that the optimal threshold increases as the weight increases, indicating the influence of the class weight on the optimal threshold. As the weight $w_{c_1}$ increases, the class-1 users contribute more heavily to the total weighted load of the system. In this case, low optimal threshold value results in unstable resource allocation. Therefore, a larger threshold is required to maintain a stable allocation process across all the users. The theoretical and simulation curves overlap exactly, indicating that the optimal threshold obtained from the simulation matches accurately with that predicted by the theoretical analysis. 

\begin{figure}[htbp]
\centerline{\includegraphics[scale = 0.5]{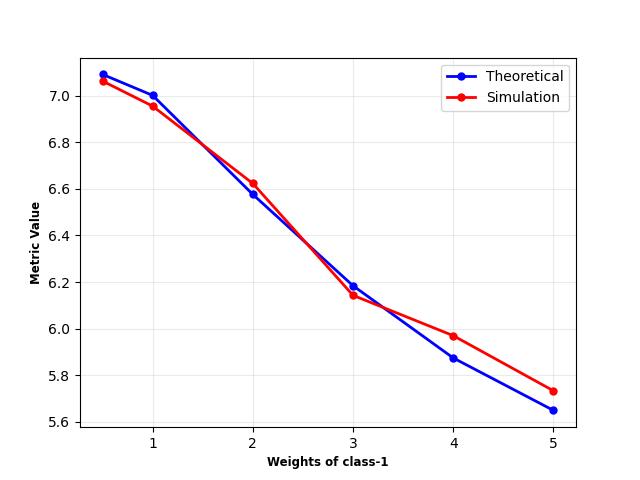}}
\caption{Comparison of overall metric value for different class-1 weights \\ under theoretical and simulation analysis}
\label{fig:Fig_05_Avg_Double_Metric_Comp}
\end{figure}
Figure~\ref{fig:Fig_05_Avg_Double_Metric_Comp} compares the theoretical and simulated values of the overall performance metric for different class-1 weight values. From the figure, it can be observed that the performance metric gradually decreases as the  weight $w_{c_1}$ increases. This behavior can be explained by the fact that increasing $w_{c_1}$  biases the resource allocation mechanism more toward class-1 users. As a result, a larger portion of the resource is allocated to the class-1 users, leaving relatively fewer resources for the class-2 users. Since the overall performance metric is computed as the weighted average across both classes, excessive prioritization of one class can lead to reduction in the aggregate performance measure. Furthermore, across the entire range of weight values, the simulated performance metric closely matches the theoretical values, with only negligible deviation. This confirms the accuracy of the mathematical formulation and numerical implementation.

\subsubsection*{Comparison of computational Time}
The theoretical analysis significantly reduces the computational time required to compute the overall performance metrics for different threshold values and determine the optimal threshold for a given system, compared with the simulation-based analysis. It is observed that, in single-class case, the theoretical analysis is 110 times faster than the corresponding simulation analysis. In double-class case, the analysis is performed for different class-1 weight values and threshold values, where the theoretical analysis is 30 times faster than the corresponding simulation analysis.

\section{Bystander Approach}
\label{sec:Bystander_approach}
The bystander approach simplifies the analysis by introducing a hypothetical user who remains in the system at all times and  evaluating the performance metrics through the time-averaged resource allocation and prospect-based outcomes experienced by this bystander. This approach is an approximation of the exact theoretical framework because the bystander is always present, hence all state transitions are conditioned on this fixed user. In contrast, an actual arriving user has a finite sojourn time and the distribution of the system observed upon arrival corresponds to the stationary distribution of other users. The performance metric is calculated based on the resource allocated to the bystander over the duration of bystander's stay in the system.

\subsubsection*{Single-class}
Let $q$ denote the total number of users in the system including the bystander. The probability that there are $q-1$ users in the $M/M/\infty$ queue is given by the distribution $\pi(q-1)$ distribution. The expected value of the time-averaged allocation of resources $\mathbb{E}[A(F)]$ is given by
\begin{equation}
\mathbb{E}[A(F)]
= \sum_{q=1}^{\infty} \pi(q-1)\,F(q)
\end{equation}

Using the stationarity of the $M/M/\infty$ queue, the time average reduces to a stationary expectation. The expected prospect contribution, $\mathbb{E}[P(F)]$ is determined by  calculating the expected number of times the transitions $q \rightarrow q \pm 1$ occur during a time interval of $T$. Based on properties of $M/M/\infty$ queues, the average number
of times $q \rightarrow q+1$ transition takes place in time $T$ is $\lambda \times \pi(q-1)$, whereas the average number of times a $q \rightarrow q-1$ transition takes place in time $T$ is $\mu \times (q-1) \times \pi(q-1)$. Hence, $\mathbb{E}[P(F)]$ is given by

\begin{equation}
\begin{aligned}
\mathbb{E}[P(F)] &=  \beta \left(
  \sum_{q=2}^{\infty} \pi(q-1) \times (q-1) \times \mu\, 
  \Delta(q-1)
\right) \\
& -\alpha \left(
  \sum_{q=1}^{\infty} \pi(q-1) \times \lambda \times
  \Delta(q+1)
\right)
\end{aligned}
\end{equation}

Where $\Delta(q+1) = F(q) - F(q+1)$ and $\Delta(q-1) = F(q-1) - F(q)$ denote the change in the allocated resources when a user arrives at or departs from the system, respectively. The overall performance metric $ E[M(F)]$ is obtained by combining the expected time-average resource allocation and the expected prospect contribution.
\begin{equation}
    \mathbb{E}[M(F)] = \mathbb{E}[A(F)] + \gamma \,\mathbb{E}[P(F)]
\end{equation}

\begin{figure}[htbp]
\centerline{\includegraphics[scale = 0.5]{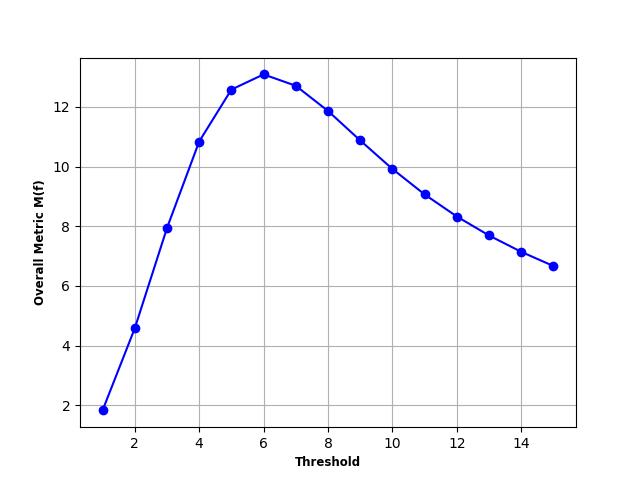}}
\caption{Overall performance metric  for single-class using bystander approach}
\label{fig:Fig_06_Bystander_single}
\end{figure}

To evaluate the performance of this approach, the system is modeled using an $M/M/\infty$ queue with fixed arrival and service rates and a total available resource of 100 units is considered for allocation to the bystander. The parameters associated with the prospect term are kept identical to the exact analytical approach, so that the impact of the prospect term can be analyzed consistently. The overall metric is evaluated over a threshold ranged of 1 to 16 and the result is shown in Fig.~\ref{fig:Fig_06_Bystander_single}.

Similar to the exact theoretical framework, the overall performance metric is low at smaller threshold values due to the fluctuations in the allocated resources. As threshold increases, the performance metric increases and reaches its maximum at the threshold $t = 6$. This corresponds to the optimal threshold, where the system allocates the resources more efficiently. Beyond this point, the performance metric gradually decreases as the threshold continues to increase.

\subsubsection*{Double-class}
The bystander approach is extended to the double-class setting, where users belong to either class-1 or class-2. Let $(q_{c_1}, q_{c_2})$ denote the total numbers of class-1 and class-2 users in the system, including the bystander. Let $\pi_{c_1}(\cdot)$ and $\pi_{c_2}(\cdot)$ denote the stationary distributions of the number of class-1 and class-2 users excluding the bystander in the corresponding $M/M/\infty$ queues.

When the bystander belongs to class-1, the system state $(q_{c_1}, q_{c_2})$ corresponds to $(q_{c_1}-1, q_{c_2})$ other users. Similarly, when the bystander belongs to class-2, the system state corresponds to $(q_{c_1}, q_{c_2}-1)$ other users. Let $G_{c_1}(q_{c_1},q_{c_2})$ and $G_{c_2}(q_{c_1},q_{c_2})$ denote the resource allocation to a class-1 and class-2 bystander respectively.

The expected time-averaged resource allocated is given by
\begin{equation}
\begin{aligned}
E[A(G)] &= 
 p_b^{(c_1)} \sum_{\substack{q_{c_1}=1 \\ q_{c_2}=0}}^{\infty} 
\pi_{c_1}(q_{c_1}-1)\,\pi_{c_2}(q_{c_2})\,
G_{c_1}(q_{c_1},q_{c_2}) \\
 &+ p_b^{(c_2)} \sum_{\substack{q_{c_1}=0\\q_{c_2}=1}}^{\infty} 
\pi_{c_1}(q_{c_1})\,\pi_{c_2}(q_{c_2}-1)\,
G_{c_2}(q_{c_1},q_{c_2})
\end{aligned}
\end{equation}

where $p_b^{(q_{c_1})}$ and $p_b^{(q_{c_2})}$ denote the probabilities that the bystander belongs to class-1 and class-2 respectively. Using the stationarity of the $M/M/\infty$ queues, the time average again reduces to a stationary expectation.

Analogous to the single-class case, the expected allocation change rates due to arrivals and departures of other users are accounted for determining $\mathbb{E}[P(G)]$. For a class-1 bystander in state $(q_{c_1}, q_{c_1})$, the arrivals of class-1 and class-2 users occur at rates $\lambda_{c_1}$ and $\lambda_{c_2}$ respectively, each contributing a loss proportional to the decrease in resource allocation. The departures of other class-1 and class-2 users occur at rates $(q_{c_1}-1)\mu$ and $q_{c_2}\mu$, producing gains due to increase in resource allocation. Combining these contributions yields the expected prospect contribution for a class-1 bystander. An analogous expression holds for a class-2 bystander, with departures occurring at rates $q_{c_1}\mu$ and $(q_{c_2}-1)\mu$. Hence, the expected prospect term is given by

\begin{equation}
\begin{aligned}
\mathbb{E}[P(G)] &=
p_b^{(c_1)}
\sum_{\substack{q_{c_1}=1\\q_{c_2}=0}}^{\infty} 
\pi_{c_1}(q_{c_1}-1)\,\pi_{c_2}(q_{c_2})
\\
& \Big(-\alpha_{c_1} \Delta_{c_1, +1}(q_{c_1},q_{c_2}) + \beta_{c_1} \Delta_{c_1, -1}(q_{c_1}, q_{c_2})\Big)\\
&+
p_b^{(c_2)}
\sum_{\substack{q_{c_1}= 0\\q_{c_2}=1}}^{\infty} 
\pi_{c_1}(q_{c_1})\,\pi_{c_2}(q_{c_2}-1) 
\\
&\Big(-\alpha_{c_2} \Delta_{c_2, +1}(q_{c_1},q_{c_2}) + \beta_{c_2} \Delta_{c_2, -1} (q_{c_1}, q_{c_2})\Big)
\end{aligned}
\end{equation}

where $\Delta_{c_1, +1}(q_{c_1},q_{c_2})$ and $\Delta_{c_2, +1}(q_{c_1},q_{c_2})$ denote the changes in the allocated resources when a class-1 and class-2 user enters the system respectively. Similarly, the changes in the allocated resources when a class-1 and class-2 user departs from the system are denoted by $\Delta_{c_1, -1}(q_{c_1},q_{c_2})$ and $\Delta_{c_2, -1}(q_{c_1},q_{c_2})$ respectively.

\begin{equation}
\begin{aligned}
\Delta_{c_1, +1}(q_{c_1},q_{c_2})
&=\lambda_{c_1}
\Bigl(
G_{c_1}(q_{c_1},q_{c_2}) - G_{c_1}(q_{c_1}+1,q_{c_2}) 
\Bigr)\\
&+ \lambda_{c_2}
\Bigl(
G_{c_1}(q_{c_1},q_{c_2}) - G_{c_1}(q_{c_1},q_{c_2}+1)
\Bigr)
\nonumber
\end{aligned}
\end{equation}

\begin{equation}
\begin{aligned}
\Delta_{c_1, -1}&(q_{c_1},q_{c_2})
=q_{c_2} \mu
\Bigl(G_{c_1}(q_{c_1},q_{c_2}-1) - G_{c_1}(q_{c_1},q_{c_2})
\Bigr)\\
&+(q_{c_1}-1)\mu
\Bigl( G_{c_1}(q_{c_1}-1,q_{c_2}) - G_{c_1}(q_{c_1},q_{c_2})
\Bigr)
\nonumber
\end{aligned}
\end{equation}

\begin{equation}
\begin{aligned}
\Delta_{c_2, +1}(q_{c_1},q_{c_2})
&=
\lambda_{c_1}
\Bigl(
G_{c_2}(q_{c_1},q_{c_2}) - G_{c_2}(q_{c_1}+1,q_{c_2}) 
\Bigr)
\\
&+
\lambda_{c_2}
\Bigl(
G_{c_2}(q_{c_1},q_{c_2}) - G_{c_2}(q_{c_1},q_{c_2}+1)
\Bigr)
\nonumber
\end{aligned}
\end{equation}

\begin{equation}
\begin{aligned}
\Delta_{c_2, -1}&(q_{c_1},q_{c_2})
=
q_{c_1}\mu
\Bigl( G_{c_2}(q_{c_1}-1,q_{c_2}) - G_{c_2}(q_{c_1},q_{c_2})
\Bigr)
\\
&+
(q_{c_2}-1) \mu
\Bigl(G_{c_2}(q_{c_1},q_{c_2}-1) - G_{c_2}(q_{c_1},q_{c_2})
\Bigr)
\nonumber
\end{aligned}
\end{equation}

The overall performance metric is given by
\begin{equation}
\mathbb{E}[M(G)] = \mathbb{E}[A(G)] + \gamma\, \mathbb{E}[P(G)].
\end{equation}

This expression depends only on the threshold and the relative weights. 

\begin{figure}[htbp]
\centerline{\includegraphics[scale = 0.55]{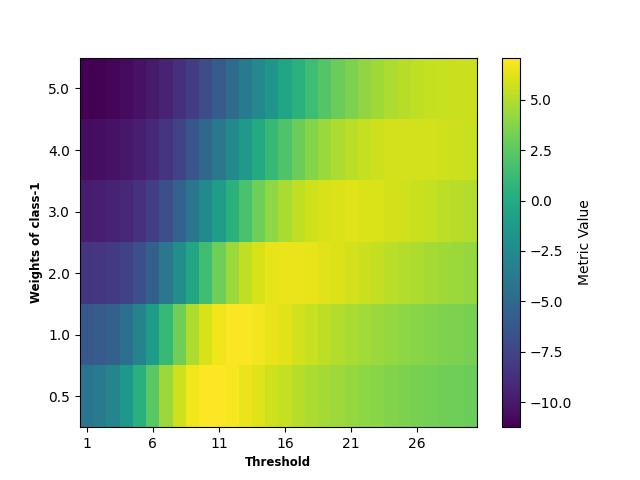}}
\caption{Heatmap of performance metric across threshold  and class-1 \\ weight values under bystander approach}
\label{fig:Fig_07_Bystander_Double_Th_Heatmap}
\end{figure}

The bystander approach for double-class case is evaluated by considering the same system settings as those used in the exact theoretical framework to ensure consistency in evaluation. The system consists of two classes with arrival rates $\lambda_{c_1} = 2$ and $\lambda_{c_2} = 6$  and a common service parameter $\mu = 1$. The same prospect metric parameters are also used. The overall performance metric is evaluated for varying class-1 weight values and threshold values. The heat map for the overall performance metric over varying class-1 weight values and threshold values is shown in Fig.~\ref{fig:Fig_07_Bystander_Double_Th_Heatmap}.

Similarly to the heatmaps obtained using the exact theoretical analysis, the darker region in the heatmap indicates that the overall performance metric is very low at lower threshold values. This is due to the frequent fluctuations in the allocated resources that occur when the number of users exceeds the threshold. Then, for all weight values, the performance metric gradually increases as the threshold increases and reaches a peak. This corresponds to the optimal threshold, where the system allocates the resources in a balanced and efficient way. Beyond the optimal threshold, the performance metric gradually decreases. This is due to the conservative allocation of the resources at high threshold values. Furthermore, as the class-1 weight increases, the peak shifts to the right, indicating that a higher threshold is required to maximize the performance metric when class-1 users are assigned greater importance. From the heatmap, the optimal threshold values can be identified for achieving better resource allocation at different weight values.

\section{Comparative Performance Evaluation between two approaches}
\label{sec:comparative_performance_evaluation}

\subsection*{Single-class} 
\begin{figure}[htbp]
\centerline{\includegraphics[scale = 0.55]{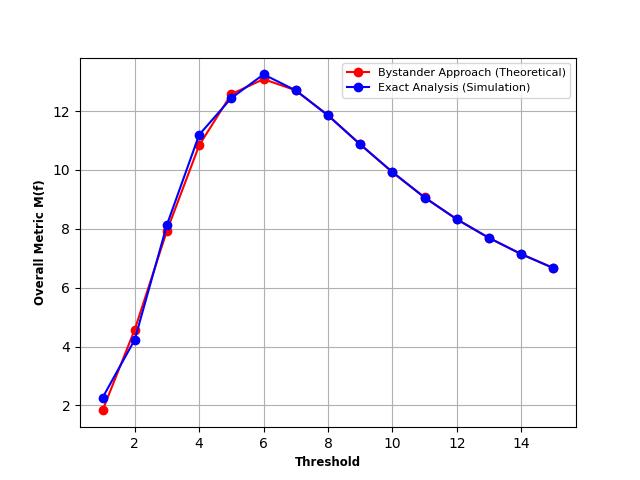}}
\caption{Performance comparison between the two approaches}
\label{fig:Fig_New_Comp_Approach_Singlepack}
\end{figure}

Figure~\ref{fig:Fig_New_Comp_Approach_Singlepack} presents the  performance comparison between the bystander approach and the exact theoretical analysis for determining the optimal threshold that maximizes the expected performance metric $E[M(F)]$. In both approaches, the optimal threshold is obtained using a uni-modal binary search over the same parameter range. For the system with parameters $\lambda = 4$, $\mu = 1$, $B = 100$, $\alpha = 2$, $\beta = 1$ and $\gamma = 1$, both approaches produce the same optimal threshold, ${t} = 6$ with the an expected overall performance metric value $\mathbb{E}[M(F)] = 13.085506.$

The agreement between the results confirms that both formulations compute the same performance metric despite their different analytical structures. Although both methods produce the same metric and the same optimal threshold, their execution time differ noticeably. The bystander approach is approximately 35 times faster than the exact theoretical analysis, which is approximately 110 times faster than the corresponding simulation-based analysis. Therefore, the bystander approach offers a substantial computational advantage, despite both approaches yielding identical optimal results.

\subsection*{Double-class}
The computational performance comparison is carried out for the double-class setting, where the optimal threshold $t$ is determined using uni-modal binary search for the system with parameters $\lambda_{c_1} = 2$, $\lambda_{c_2} = 6$, $\mu = 1$, $B = 100$,  $\alpha_{c_1} = 4$, $\alpha_{c_2} = 3$, $\beta_{c_1} = 1$, $\beta_{c_2} = 2$, $\gamma = 1$ and weights $w_{c_1} = 3$ and $w_{c_2} = 1 $. Both approaches yield the same optimal threshold ${t} = 21$. Similar to the single-class case, this confirms the consistency of the two approaches despite their different analytical structures. Furthermore, the bystander approach is approximately 10 times faster than exact theoretical analysis, which is approximately 30 times faster than the corresponding simulation-based analysis in double-class scenario. 

\begin{figure}[htbp]
\centerline{\includegraphics[scale = 0.55]{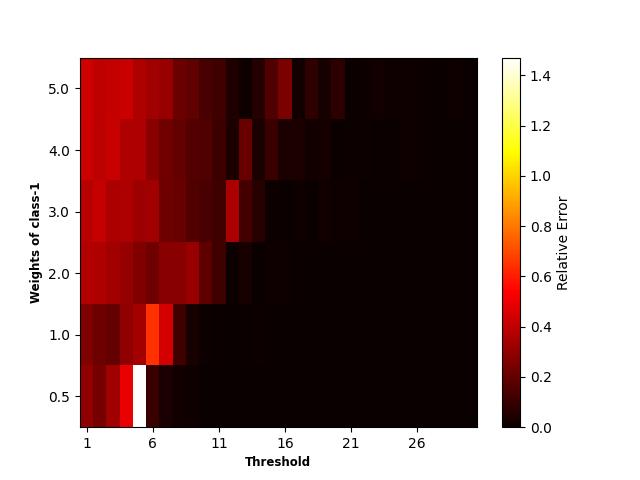}}
\caption{Performance comparison between the two approaches}
\label{fig:Fig_New_Relative_Error}
\end{figure}

Furthermore, the two approaches are compared for different class-1 weight values and the relative error between the performance metrics obtained using the two approaches is determined. The heatmap of the relative error is shown in Fig.~\ref{fig:Fig_New_Relative_Error}. From the heatmap, the relative error is slightly higher at smaller threshold values, particularly when the class-1 weight is low, as indicated by brighter regions. As the threshold increases, the relative error decreases significantly, as reflected by the darker regions. Overall, the heatmap demonstrates that both approaches produce similar overall performance metric values with only slight deviations.

\section{Conclusion}
\label{sec:Conclusion}

Motivated by decades of highly influential work in behavioral economics, we introduce a general utility metric for user satisfaction during multimedia streaming, which captures the asymmetric humans response to the same amount of increase and decrease in service quality. We propose a threshold-based resource allocation policy that reduces both the frequency and the amount of drops in allocated resources and, consequently, in service quality, while respecting the priorities of different classes of users. Through extensive simulations, we observe that, for a well chosen threshold, which may depend on the network configuration, this policy outperforms the existing fair allocation policy. Building on the detailed queuing theoretic analysis of the system, we propose a fast method for determining the exact optimal threshold  and further propose an even faster method to find the threshold approximately. These methods are of practical importance since the network scenarios change dynamically in practice, thereby requiring the optimal threshold to be recalculated frequently. 

\bibliographystyle{IEEEtran}
\bibliography{references}

\end{document}